\documentclass[12pt]{article}
\pdfoutput=1

\usepackage{jheppub}
\usepackage{amssymb,amsfonts}
\usepackage{epsfig}
\usepackage{tikz}
\usepackage{enumerate}
\usepackage{mathrsfs}
\usepackage{graphicx}
\usepackage{latexsym}
\usepackage{enumerate}

\newcommand{\be}{\begin{equation}}
\newcommand{\ee}{\end{equation}}
\newcommand{\bea}{\begin{eqnarray}}
\newcommand{\eea}{\end{eqnarray}}

\newcommand{\lr}{\left (}
\newcommand{\rr}{\right )}
\newcommand{\ls}{\left [}
\newcommand{\rs}{\right ]}

\newcommand{\lambdal}{\bar\lambda}
\newcommand{\lambdar}{\lambda}
\newcommand{\zetar}{\eta}
\newcommand{\tl}{\tau}
\newcommand{\tlV}{\tau}
\newcommand{\ttr}{H}
\newcommand{\ttrV}{E}
\newcommand{\tB}{B}
\newcommand{\wm}{h}
\newcommand{\wmV}{e}

\newcommand{\p}{\partial}

\renewcommand{\bar}[1]{\overline{#1}}
\renewcommand{\tilde}[1]{\widetilde{#1}}

\makeatletter
\renewcommand{\@seccntformat}[1]{\csname the#1\endcsname.\,\,}
\makeatother

\let \savenumberline \numberline
\def \numberline#1{\savenumberline{#1.}}

\makeatletter
\def\@fpheader{\relax}
\makeatother

\title{\ \vspace{1.6cm} \\
Nonrelativistic String Theory and T-Duality}
\author{Eric Bergshoeff${}^{\, a}$, Jaume Gomis${}^{\, b}$, and Ziqi Yan${}^{\, b}$}
\emailAdd{e.a.bergshoeff@rug.nl}
\emailAdd{jgomis@pitp.ca}
\emailAdd{zyan@pitp.ca}
\affiliation{
${}^a$Van Swinderen Institute, University of Groningen\\
Nijenborgh 4, 9747 AG Groningen, The Netherlands\medskip\\
${}^b$Perimeter Institute for Theoretical Physics\\
31 Caroline St N, Waterloo, ON N2L 6B9, Canada}
\abstract{Nonrelativistic string theory in flat spacetime is described by a two-dimensional quantum field theory with a nonrelativistic global symmetry acting on the worldsheet fields. Nonrelativistic string theory is   unitary,   ultraviolet complete
 and has a string spectrum and spacetime S-matrix enjoying nonrelativistic   symmetry.
The worldsheet theory of nonrelativistic string theory is coupled to a curved spacetime background and to a Kalb-Ramond two-form and   dilaton field. The appropriate spacetime geometry for nonrelativistic string theory is 
dubbed string Newton-Cartan geometry,  which  is distinct from Riemannian geometry.
 This  defines the sigma model of nonrelativistic string theory describing strings propagating and interacting in   curved background fields. We also implement    T-duality transformations in the path integral of this  sigma model   and uncover the spacetime interpretation of T-duality. We show that T-duality along the longitudinal direction of the string Newton-Cartan geometry  describes relativistic string theory on a  Lorentzian geometry with a compact lightlike isometry, which is otherwise only defined by a subtle infinite boost limit. This relation provides a first principles definition of string theory in the
discrete light cone quantization (DLCQ) in an arbitrary background, a quantization that appears in nonperturbative approaches to quantum field theory and string/M-theory, such as in Matrix theory. T-duality along a transverse direction of the string Newton-Cartan geometry equates nonrelativistic string theory in two distinct, T-dual backgrounds.}

\begin{document}
\maketitle

\section{Introduction}

A beautiful feature of  string theory is the intricate interplay between worldsheet and target space physics. The global symmetries of the two-dimensional quantum field theory (QFT) on the string worldsheet  encode the   symmetries of the target space   geometry. Vertex operators of the two-dimensional QFT correspond to physical excitations propagating in the target space background, and correlation functions of the worldsheet theory determine the spacetime S-matrix.

A striking and originally unwarranted  prediction of  string theory is the existence of a vertex operator  corresponding to a  massless spin two excitation in the target space. This excitation has the quantum numbers of the quantum of geometry, the graviton. The low energy tree-level S-matrix of string theory  around Minkowski spacetime  is that of General Relativity, which unavoidably emerges from the dynamics  of   relativistic string theory.

In \cite{Gomis:2000bd} a consistent,  unitary and ultraviolet complete string theory described by a  two-dimensional QFT with a  (string)-Galilean invariant global symmetry  was put forward. This string theory has additional   worldsheet fields beyond those parametrizing spacetime coordinates. These additional fields  
play a central role for the consistency of this string theory.\,\footnote{The construction  in  \cite{Gomis:2000bd} was motivated in part by  \cite{Klebanov:2000pp}. See also \cite{Danielsson:2000gi}.} This  novel type of string theory was dubbed nonrelativistic string theory \cite{Gomis:2000bd}.\,\footnote{In order to avoid potential confusions, we emphasize that the two-dimensional QFT is relativistic   and that the nonrelativistic symmetries act on the target space, i.e.~on the worldsheet fields. Nonrelativistic string theory is defined by a sum over two-dimensional Riemann surfaces. The special structure of the worldsheet theory localizes the path integral of nonrelativistic string theory to submanifolds in the moduli space of Riemann surfaces (see
\cite{Gomis:2000bd} for details).}  This string theory was shown to be endowed with a spectrum of string excitations with a
(string)-Galilean invariant dispersion relation and S-matrix. Nonrelativistic string theory has a simple target space interpretation: it describes strings propagating and interacting in   a   string-Galilean invariant flat spacetime background geometry \cite{Gomis:2000bd}. The target space  geometry of   nonrelativistic string theory differs from the conventional Riemannian one, in particular there is no    Riemannian, Lorentzian metric in the target space. Likewise, the spacetime effective action of nonrelativistic string theory is not described at low energies by General Relativity. Indeed,    nonrelativistic string theory does not have   massless particles and  is therefore not described at low energies by General Relativity. Nonrelativistic string theory,  being ultraviolet finite,  provides a quantization of nonrelativistic spacetime geometry akin to how relativistic string theory provides a quantization of Riemannian geometry and  of (Einstein) gravity.

We couple nonrelativistic string theory     to background fields: a curved target space geometry, a Kalb-Ramond two-form field  and a dilaton. This defines the  nonlinear sigma model   describing string propagation on a nonrelativistic target space structure with background fields, which we will write down in this paper.\,\footnote{See also \cite{Gomis:2005pg}.}
The appropriate spacetime geometry that the nonrelativistic string couples to is the so-called string Newton-Cartan geometry \cite{stringyNC,stringyNClimit}, a geometric structure that is distinct from a Riemannian metric.\,\footnote{We emphasize that this is also different from the well-studied Newton-Cartan geometry (more below). For other recent work on strings propagating in different nonrelativistic backgrounds, see \cite{Batlle:2016iel,Gomis:2016zur,Batlle:2017cfa,HHO,Kluson}. Also see footnote \ref{footnote:TNC} for a more precise relation between \cite{HHO} and the string Newton-Cartan geometry.}
Quantum consistency of the  nonlinear sigma model determines the   background fields on which    nonrelativistic string theory can be consistently  defined.  Nonrelativistic string theory provides a quantum definition of the classical target space theory that appears in the low energy expansion.

In this work we also  study  T-duality of the path integral defining nonrelativistic string theory on an arbitrary string Newton-Cartan spacetime background and in the presence of a Kalb-Ramond and dilaton field.
The string Newton-Cartan spacetime geometry of nonrelativistic string theory admits   two physically distinct T-duality transformations: longitudinal and transverse. This is a consequence of the foliation of the string Newton-Cartan
structure that the nonrelativistic string couples to. We derive the explicit form of the T-dual background fields in
nonrelativistic string theory.

An interesting  conclusion is reached in the study of   longitudinal T-duality.
We show that T-duality along a longitudinal spatial direction leads to a worldsheet theory that admits the following interesting interpretation: it is the worldsheet theory of a relativistic string propagating on a Riemannian, Lorentzian manifold with a compact lightlike isometry and in the presence of Kalb-Ramond and dilaton fields!\,\footnote{See also \cite{Gomis:2000bd,Danielsson:2000gi,Gomis:2005pg}.}
Therefore,  nonrelativistic string theory on a string Newton-Cartan geometry with a longitudinal isometry can be used to solve for the quantum dynamics of   relativistic string  theory on a Riemannian, Lorentzian manifold with a compact lightlike isometry in the discrete light cone quantization (DLCQ).
The DLCQ of QFTs and string/M-theory plays an important role in nonperturbative approaches to QCD and in Matrix theory~\cite{Banks:1996vh,Susskind:1997cw,Seiberg:1997ad,Sen:1997we}.
Previously, the DLCQ of string theory was only defined via a subtle limit of compactification on a spacelike circle \cite{Seiberg:1997ad,Sen:1997we,Hellerman:1997yu}. Instead, we find that the relation to nonrelativistic string theory via a longitudinal T-duality transformation  provides a first principles definition of string theory in the DLCQ on arbitrary Lorentzian backgrounds with a lightlike isometry. The DLCQ   of relativistic string theory on a Lorentzian geometry is thus described by the  sigma model of nonrelativistic string theory, with additional worldsheet fields beyond those corresponding to spacetime coordinates.

For the convenience of the reader, we summarize here the results of performing the   T-duality transformation of nonrelativistic string theory according to the nature of the isometry direction:

\begin{enumerate}
	
\item

\textbf{Longitudinal spatial T-duality}:
Nonrelativistic string theory   on a  string Newton-Cartan background is mapped to relativistic string theory on a  Riemannian, Lorentzian background geometry with a compact lightlike isometry. See \S\ref{sec:longflat} for the precise mapping between  the string Newton-Cartan data  with  background    Kalb-Ramond and dilaton fields, and the Lorentzian metric with background Kalb-Ramond and dilaton fields.

\item

\textbf{Longitudinal lightlike T-duality}:
Nonrelativistic string theory   on a  string Newton-Cartan background is mapped to nonrelativistic string theory   on a  T-dual string Newton-Cartan background with a longitudinal lightlike isometry.  The precise mapping between  the two T-dual string Newton-Cartan background fields can be found in \S\ref{sec:null}.

\item

\textbf{Transverse   T-duality}:
Nonrelativistic string theory   on a  string Newton-Cartan background is mapped to nonrelativistic string theory   on a  T-dual string Newton-Cartan background. See \S\ref{sec:trans} for the precise T-duality transformation rules.

\end{enumerate}

The plan for the remainder of this paper is as follows. In \S\ref{sec:NCstring} we describe the string Newton-Cartan geometry  that   nonrelativistic string theory can be coupled to. We proceed to write down the sigma model describing nonrelativistic string theory coupled to such a  string Newton-Cartan background, together with   a Kalb-Ramond two-form  field and a dilaton. We study the path integral of this sigma model and  study T-duality along a longitudinal spatial direction in \S\ref{sec:longflat}, a longitudinal lightlike direction in \S\ref{sec:null} and a transverse spatial direction in \S\ref{sec:trans}. Finally, in \S\ref{sec:concl} we present our conclusions.

\section{Nonrelativistic String Theory in a String Newton-Cartan Background} \label{sec:NCstring}

In this section we present the  construction of the two-dimensional nonlinear sigma model describing nonrelativistic  string theory on a  string Newton-Cartan background   in the presence of a Kalb-Ramond two-form  field and a dilaton (see also \cite{Gomis:2005pg,stringyNC}).
 This sigma model  extends the worldsheet theory in  flat spacetime of  \cite{Gomis:2000bd} to arbitrary curved background fields. In  \S\ref{sec:sNCg}  we review some basic properties of this  string Newton-Cartan background spacetime structure.\,\footnote{The corresponding spacetime nonrelativistic gravity theory  was called ``stringy" Newton-Cartan gravity in \cite{stringyNC}. An extensive description improving  a few results of \cite{stringyNC} can be found in \cite{stringyNClimit}.} Subsequently, in \S\ref{sec:NCs}, we discuss the nonrelativistic string sigma model action coupled to this   geometry and background fields.

\subsection{String Newton-Cartan  Geometry} \label{sec:sNCg}

We  define  string Newton-Cartan geometry on a $D+1$ dimensional spacetime manifold $\mathcal{M}$ as follows. Let $\mathcal{T}_p$ be the tangent space attached to a point $p$ in $\mathcal{M}$. We decompose $\mathcal{T}_p$ into two \emph{longitudinal} directions indexed by $A = 0, 1$ and $D-1$ \emph{transverse} directions indexed by $A' = 2, \cdots, D$, respectively.\,\footnote{A particular  curved spacetime foliation  structure of string Newton-Cartan type   appeared in \cite{Bagchi:2009my} as the outcome of  the nonrelativistic limit of string theory on $AdS_5\times S^5$ \cite{Gomis:2005pg}.} A two-dimensional foliation is attributed to $\mathcal{M}$ by introducing a generalized clock function $\tlV_\mu{}^A$, also called the longitudinal Vielbein field,  that satisfies the constraint
\begin{equation}\label{foliation}
D_{[\mu}\tlV_{\nu]}{}^A=0\,.
\end{equation}
The derivative $D_\mu$ is covariant with respect to the longitudinal Lorentz transformations acting on the index $A$.\,\footnote{$D_\mu$ contains a dependent spin-connection field $\omega_\mu{}^{AB}(\tl)$ whose explicit expression will not be needed here. For more details, see \cite{stringyNC,stringyNClimit}.} As a consequence of the foliation constraint \eqref{foliation}, we have
\begin{equation}
\partial_{[\mu}\big(\tlV_{\nu\phantom{]}\!}{}^A\tlV_{\rho]}{}^B\epsilon_{AB}\big) =0\hskip 1truecm  \rightarrow \hskip 1truecm
\tlV_\mu{}^A\tlV_\nu{}^B\epsilon_{AB} = \partial_{[\mu}\rho_{\nu]}
\end{equation}
for some vector field $\rho_\mu$.

We consider now the following transformations with corresponding generators:
\begin{subequations}
\begin{align*}
	\text{longitudinal translations} & \qquad H_A \\
	\text{transverse translations} & \qquad P_{A'} \\
	\text{string Galilei boosts} & \qquad G_{AA'} \\
	\text{longitudinal Lorentz rotations} & \qquad M_{AB} \\
	\text{transverse spatial rotations} & \qquad J_{A'B'}
\end{align*}
\end{subequations}
We   refer to the Lie algebra spanned by these generators as  the \emph{string Galilei algebra} \cite{Gomis:2000bd, nrGalilei, stringyNC, NCbranes}. This defines the local spacetime symmetry   that replaces the spacetime Lorentz symmetry $SO(D,1)$  in the relativistic case.
Besides the longitudinal Vielbein field $\tlV_\mu{}^A$ corresponding to $H_A$,  we only introduce the transverse Vielbein field $\ttrV_\mu{}^{A'}$ corresponding to the generators $P_{A^\prime}$. The dependent spin-connection fields corresponding to the other generators $G_{AA'}\,, M_{AB}$ and $J_{A'B'}$  will not be needed in what follows.

 The (projective) inverse Vielbein fields $\tlV^\mu{}_A$ and $\ttrV^\mu{}_{A^\prime}$ corresponding to $\tlV_\mu{}^A$ and $\ttrV_\mu{}^{A^\prime}$, respectively,  are defined via the relations
\begin{subequations}
\begin{align}
	\ttrV_\mu{}^{A'} \ttrV^\mu{}_{B'} & = \delta^{A'}_{B'}\,,
		&
	\tlV^\mu{}_A \tlV_\mu{}^B & = \delta^B_A\,,
		&
	\tlV_\mu{}^A \tlV^\nu{}_A + \ttrV_\mu{}^{A'} \ttrV^\nu{}_{A'} & = \delta^\nu_\mu\,, \\[.2truecm]
	\tlV^\mu{}_A \ttrV_\mu{}^{A'} & = 0\,,
	 	&
	\tlV_\mu{}^A \ttrV^\mu{}_{A'} & = 0\,.
\end{align}
\end{subequations}
Parametrizing the string Galilei boost transformations by $\Sigma_A{}^{A'}$, the Vielbeine and their inverses transform under string Galilei boosts as follows:
\begin{subequations} \label{eq:boosts}
\begin{align}
	\delta^{}_\Sigma \tlV_\mu{}^A & = 0\,,
&
	\delta{}_\Sigma \ttrV_\mu{}^{A'} &= - \tlV_\mu{}^A\Sigma_A{}^{A'}\,,\\[.2truecm]
\delta^{}_\Sigma \tlV^\mu{}_A  &= \ttrV^\mu{}_{A^\prime}\Sigma_A{}^{A^\prime}\,,
&
	\delta{}_\Sigma \ttrV^\mu{}_{A'} & = 0\,.
\end{align}
\end{subequations}
From the Vielbeine we construct a longitudinal metric $\tl_{\mu\nu}$ and a transverse metric $\ttr^{\mu\nu}$,
\be
	\tl_{\mu\nu} \equiv \tlV_\mu{}^A \tlV_\nu{}^B \eta_{AB}\,,\hskip 2truecm
	\ttr^{\mu\nu} \equiv \ttrV^\mu{}_{A'} \ttrV^\nu{}_{B'} \delta^{A'B'}\,.
	\label{longinvmetr}
\ee
Both metrics are not only invariant under the (longitudinal and transverse) rotations but also invariant under the string Galilei boost transformations \eqref{eq:boosts}. They are orthogonal in the sense that $\tl_{\mu\rho}\ttr^{\rho\nu}=0$.

In order to write down the action for a string moving in a  string Newton-Cartan background, we will also need a  transverse two-tensor $\ttr_{\mu\nu}$ with covariant indices.\,\footnote{A longitudinal two-tensor $\tl^{\mu\nu}$ with contra-variant indices will not be needed.} However, the na\"{i}ve choice, $\ttrV_\mu{}^{A'} \ttrV_\nu{}^{B'} \delta_{A'B'}$, is not invariant under the string Galilei boosts  \eqref{eq:boosts}. The lack of a boost-invariant inverse for $\ttr^{\mu\nu}$ (and similarly for $\tl_{\mu\nu}$) prohibits the longitudinal and transverse metrics from combining into a single Riemannian metric on $\mathcal{M}$.

Constructing a  boost-invariant  transverse two-tensor $\ttr_{\mu\nu}$  requires introducing a noncentral extension $Z_A$ of the string Galilei algebra  that occurs in  the following commutation relations:\,\footnote{When $Z_A$ is included in the string Galilei algebra, requiring the Jacobi identities to hold leads to a further extension by  a generator $Z_{AB}$ with $Z_{AB} = - Z_{BA}$ \cite{nrGalilei,stringyNClimit}. The gauge field associated to this generator will not play a role in this paper.}
\be
	[G_{AA'}, P_{B'}] = \delta_{A'B'} Z_A\,.
\ee
We introduce gauge fields  $m_\mu{}^A$ corresponding to the generators $Z_A$, which transform under a gauge transformation with parameter $\sigma^A$ and under the Galilean boosts  as
\be \label{eq:deltammuA}
	\delta  m_\mu{}^A = D_\mu\sigma^A + \ttrV_\mu{}^{A^\prime} \Sigma^A{}_{A'} \,,
\ee	
where the derivative $D_\mu$ is covariant with respect to the longitudinal Lorentz rotations. By using this extra gauge field, we can define the  boost-invariant (but not $Z_A$ gauge-invariant!) two-tensor,
\be
	\ttr_{\mu\nu} \equiv \ttrV_\mu{}^{A'} \ttrV_\nu{}^{B'} \delta_{A'B'} +  \left( \tlV_{\mu}{}^A m_{\nu}{}^B +\tlV_{\nu}{}^A m_{\mu}{}^B\right)\eta_{AB}\,.
	\label{boostinvmet}
\ee

We refer to the geometry described by the fields $\tlV_\mu{}^A\,, \ttrV_\mu{}^{A^\prime}$ and $m_\mu{}^A$ as the \emph{string Newton-Cartan geometry}.\,\footnote{In contrast to string Newton-Cartan geometry, Newton-Cartan geometry is characterized by a {\sl one-dimensional} foliation with a clock function $\tlV_\mu{}^0$ satisfying  $\partial_{[\mu}\tlV_{\nu]}{}^0=0$. We denote the generators of the Galilei algebra by
$\{H\,, P_{A^\prime}\,, G_{A^\prime}\,, J_{A^\prime B^\prime}\}$ with $A^\prime = 1, \cdots ,D$\,.
In addition to the field $\tlV_\mu{}^0$, the theory also contains a transverse Vielbein field
$\ttrV_\mu{}^{A^\prime}$, associated with the spatial translation generators $P_{A^\prime}$, and  a single central charge gauge field $m_\mu$, associated with a central charge generator $Z$. This generator $Z$ appears in the  commutator of a spatial translation and a Galilean boost generator,
 \begin{equation}
[P_{A^\prime}\,, G_{B^\prime}] = \delta_{A^\prime B^\prime}Z\,.
\end{equation}
This defines the Bargmann algebra (the centrally extended Galilei algebra). Taking the nonrelativistic limit of particles and strings coupled to general relativity, one finds that, whereas strings couple to string Newton-Cartan geometry,
particles naturally  couple to Newton-Cartan geometry: it defines the background geometric structure to which nonrelativistic QFTs in flat nonrelativistic spacetime can be canonically coupled to.\label{footnote:NC}}

\subsection{Nonrelativistic String Theory Sigma Model} \label{sec:NCs}

We  proceed now to writing down the sigma model describing nonrelativistic string theory in  a general curved
string Newton-Cartan background and in the presence of a Kalb-Ramond and dilaton field. Since the  nonrelativistic string sigma model is actually relativistic on the two-dimensional worldsheet (but not on the target space), the sigma model is defined on a Riemann surface $\Sigma$.
In nonrelativistic string theory we must integrate over all Riemann surfaces  \cite{Gomis:2000bd}.

The sigma model of nonrelativistic string theory on a string Newton-Cartan background can be constructed by deforming the  worldsheet theory in flat spacetime constructed  in \cite{Gomis:2000bd} by suitable vertex operators. These acquire an elegant spacetime interpretation as spacetime fields. The worldsheet fields of nonrelativistic string theory include worldsheet scalars parametrizing the spacetime coordinates $x^\mu$ and two    one-form fields on the worldsheet, which we denote by $\lambdar$ and $\lambdal$.\,\footnote{In spite of the additional worldsheet fields, the critical dimension of nonrelativistic string
theory is either $10$ or $26$ \cite{Gomis:2000bd}.} These additional fields are required to realize the
extended string Galilei symmetry on the worldsheet theory and are responsible for interesting  peculiarities  of nonrelativistic string   perturbation theory \cite{Gomis:2000bd}.

 Let the worldsheet surface $\Sigma$ be parametrized by $\sigma^\alpha$, with $\alpha = 0, 1$. In order to write down the action  of nonrelativistic  string theory in a curved string Newton-Cartan background, we
   pullback  from the target space $\mathcal{M}$ to the worldsheet $\Sigma$   the Vielbeine $\{\tlV_\mu{}^A\,, \ttrV_\mu{}^{A^\prime}\}$ and the covariant, string Galilei boost invariant  two-tensors $\{\tl_{\mu\nu}\,, \ttr_{\mu\nu}\}$ defined in \eqref{longinvmetr} and \eqref{boostinvmet}.
Nonrelativistic string theory   also  couples to a dilaton field $\Phi$ and  a nonrelativistic Kalb-Ramond B-field $\tB_{\mu\nu}$, both of which are target space fields defined on $\mathcal{M}$.

Nonrelativistic string theory in the Polyakov formalism is endowed with an independent  worldsheet metric $\wm_{\alpha\beta} (\sigma)$. \text{We introduce  Vielbeine $\wmV_\alpha{}^a$, $a = 0, 1$ on $\Sigma$ such that}
\be
	\wm_{\alpha\beta} = \wmV_\alpha{}^a \wmV_\beta{}^b \eta_{ab}\,.
\ee
Using  light-cone coordinates for the flat index $a$ on the worldsheet tangent space,  we define locally
\be
	\wmV_\alpha   \equiv \wmV_\alpha{}^0 + \wmV_\alpha{}^1\,, \qquad 	\bar{\wmV}_\alpha   \equiv \wmV_\alpha{}^0 - \wmV_\alpha{}^1\,.
\ee
On the other hand, using light-cone coordinates for the flat index $A$ on the spacetime tangent space $\mathcal{T}_p$, we define locally
\be
	\tlV_\mu \equiv \tlV_\mu {}^0 + \tlV_\mu {}^1\,, \qquad
	\bar{\tlV}_\mu \equiv \tlV_\mu {}^0 - \tlV_\mu {}^1\,.
 \ee
The   sigma model of nonrelativistic string theory on an arbitrary string Newton-Cartan geometry, B-field and dilaton background is given by (see also \cite{Gomis:2005pg})\,
\begin{align} \label{eq:NCstring}
	S & = -\frac{T}{2} \int d^2 \sigma \ls \sqrt{-\wm} \, \wm^{\alpha\beta} \, \p_\alpha x^\mu \p_\beta x^\nu \ttr_{\mu\nu} + \epsilon^{\alpha\beta} \lr \lambdar \, \wmV_\alpha \tlV_\mu + \lambdal \, \bar \wmV_\alpha \bar \tlV_\mu \rr \p_\beta x^\mu \rs \notag \\
		& \quad - \frac{T}{2} \int d^2 \sigma \, \epsilon^{\alpha\beta} \p_\alpha x^\mu \p_\beta x^\nu \tB_{\mu\nu} + \frac{1}{4\pi} \int d^2 \sigma \sqrt{-\wm} \, R\, \Phi \,,
\end{align}
where $\wm = \det \wm_{\alpha\beta}$,  $\wm^{\alpha\beta}$ is the inverse of $\wm_{\alpha\beta}$, $R$ is the scalar curvature of $\wm_{\alpha\beta}$ and $T$ is the string tension.
The fields  $\lambdar$ and $\lambdal$ are worldsheet scalars under diffeomorphisms. It is only after imposing the conformal gauge (see below around \eqref{eq:NCstringflat}) that they become worldsheet one-forms.
 This sigma model encodes the coupling of the worldsheet to the appropriate combination of gauge fields $\tlV_\mu{}^A\,, \ttrV_\mu{}^{A^\prime}$ and $m_\mu{}^A$ defining the  string Newton-Cartan geometry. From the point of view of the two-dimensional QFT on the worldsheet, these spacetime gauge fields are coupling constants of the QFT.

\medskip
 The symmetries of the nonrelativistic sigma model \eqref{eq:NCstring} are
\begin{itemize}
\item Worldsheet diffeomorphisms:
under a change of worldsheet coordinates $\sigma'^\alpha(\sigma)$ the worldsheet fields transform as
\begin{subequations}
\begin{align}
	 \wm'_{\alpha\beta}(\sigma')&=\frac{\partial \sigma^\gamma}{\partial \sigma'^\alpha} \frac{\partial \sigma^\delta}{\partial \sigma'^\beta} \wm_{\gamma\delta}(\sigma)\,, \\[2pt]
	 \epsilon'^{\alpha\beta} (\sigma') &=\left| \frac{\partial \sigma}{\partial  \sigma'} \right|\frac{\partial \sigma'^\alpha}{\partial \sigma^\gamma} \frac{\partial \sigma'^\beta}{\partial \sigma^\delta}	\epsilon^{\gamma\delta} (\sigma)\\[2pt]
	 x'^\mu(\sigma')&= x^\mu(\sigma)\,, \\[2pt]
	 \lambdar'(\sigma')&= \lambdar(\sigma)\,, \\[2pt]
	 \lambdal'(\sigma')&= \lambdal(\sigma)\,.
\end{align}
\end{subequations}
Note that $\lambda$ and $\bar\lambda$ also transform under Lorentz transformations on the worldsheet, which is made manifest by using the light-cone notation.

\item Worldsheet Weyl invariance: under a  local Weyl transformation $w(\sigma)$ the worldsheet fields transform as
\begin{subequations}
\begin{align}
	 \wm'_{\alpha\beta}(\sigma)&= e^{2w(\sigma)} \wm_{\alpha\beta}(\sigma)\,, \\[2pt]
	 \epsilon'^{\alpha\beta} (\sigma) &=\epsilon^{\alpha\beta} (\sigma) \,, \\[2pt]
	x'^\mu(\sigma)&= x^\mu(\sigma)\,, \\[2pt]
	\lambdar'(\sigma)&= e^{-w(\sigma)}\lambdar(\sigma)\,, \\[2pt]
	\lambdal'(\sigma)&=e^{-w(\sigma)} \lambdal(\sigma)\,.
\end{align}
\end{subequations}

\item Target space reparametrizations: under a change of worldsheet variables $x^\mu(x')$ the action \eqref{eq:NCstring} transforms covariantly if $\ttr_{\mu\nu}, \tlV_\mu$ and $\bar\tlV_\mu$ transform as tensors under spacetime diffeomorphisms
\vspace{-1.5mm}
\begin{subequations}
\begin{align}
	\ttr'_{\mu\nu}(x')&=\frac{\partial x^\rho}{\partial x'^\mu} \frac{\partial x^\sigma}{\partial x'^\nu} \ttr_{\rho\sigma}(x) \,, 
	\\[2pt]
	\tlV'_\mu(x')&= \frac{\partial x^\rho}{\partial x'^\mu}\tlV_\rho(x)\,, \\[2pt]
	\bar\tlV'_\mu(x')&= \frac{\partial x^\rho}{\partial x'^\mu}\bar\tlV_\rho(x)\,,
\end{align}
\end{subequations}
as dictated by the string Newton-Cartan geometry. Moreover, the fact that $\tau_\mu$ and $\bar\tau_\mu$ also transform under the longitudinal Lorentz transformations is made manifest by using the light-cone notation.
In addition to these longitudinal Lorentz transformations, the action \eqref{eq:NCstring} is invariant under all the other tangent space transformations generated by the extended string Galilei algebra. In the case of the $Z_A$ gauge transformations parametrized by $\sigma^A$ in \eqref{eq:deltammuA}, the worldsheet fields $\lambda$ and $\bar\lambda$ transform nontrivially as follows:
\be \label{eq:deltalambdabar}
	\delta \lambda = \frac{1}{\sqrt{-h}} \, \epsilon^{\alpha\beta} \, \bar e_\alpha \p_\beta x^\mu D_\mu \bar{\sigma}\,,	
		\qquad
	\delta \bar\lambda = \frac{1}{\sqrt{-h}} \, \epsilon^{\alpha\beta} e_\alpha \p_\beta x^\mu D_\mu \sigma\,,
\ee
where $\sigma \equiv \sigma^0 + \sigma^1$ and $\bar{\sigma} \equiv \sigma^0 - \sigma^1$\,. Note that the gauge parameter $\sigma^A$ used here is not to be confused with the worldsheet coordinates $\sigma^\alpha$\,. We also note that the action \eqref{eq:NCstring} is only invariant under the $\sigma^A$ transformations when the constraint $D_{[\mu} \tau_{\nu]}{}^A = 0$ in \eqref{foliation} is imposed.\footnote{In \cite{HHO}, strings in a different nonrelativistic spacetime geometry are introduced from a rather different perspective. However, if one requires the zero torsion condition $d \tau = 0$ in \cite{HHO}, then the theory considered there can be reinterpreted as a string propagating in Newton-Cartan geometry with an additional worldsheet scalar representing the longitudinal spatial direction along the string. This geometry is a special case of string Newton-Cartan geometry (with zero Kalb-Ramond and dilaton field) and can be obtained from the general case considered in the current paper by a reduction over the longitudinal spatial direction followed by a truncation.\label{footnote:TNC}}

\end{itemize}

Imposing quantum mechanical Weyl invariance  of the path integral based on the action  \eqref{eq:NCstring}, that is setting the beta-functions of the background fields to zero, determines the spacetime background fields on which nonrelativistic string theory can be consistently defined. This parallels the   mechanism which  determines the consistent backgrounds of relativistic string theory and that leads to     Einstein's equations in relativistic string theory
\cite{Friedan:1980jm,Callan:1985ia}. In nonrelativistic string theory the consistent backgrounds are  solutions of a nonrelativistic gravitational theory \cite{stringyNClimit}.

The string Newton-Cartan background fields that describe nonrelativistic string theory in flat spacetime   are
\be
\tlV_\mu{}^{A}= \delta^A_\mu\,,\qquad \ttrV_\mu{}^{A'}=  \delta^{A'}_\mu \,,\qquad m_\mu{}^{A}=0\,.
\ee
The nonlinear sigma model \eqref{eq:NCstring} with these background fields reproduces the action of nonrelativistic string theory in flat spacetime in the conformal gauge \cite{Gomis:2000bd},
\begin{align} \label{eq:NCstringflat}
	S & = -\frac{T}{2} \int d^2 \sigma \lr  \p x^{A'} \, \bar\p x^{B'} \delta_{A'B'} +    \lambdar \, \bar \partial X + \lambdal \, \partial \bar X \rr \,,
\end{align}
where for simplicity we have set $\tB_{\mu\nu} = 0$\,. We also defined
\be
	X \equiv     x ^0 + x^1 \,,
		\qquad
	\bar X \equiv  x^0 - x^1  \,,
\ee
as well as
\be
	\p \equiv \frac{\p}{\p\sigma^0} + \frac{\p}{\p\sigma^1}\,,
		\qquad
	\bar\p \equiv - \frac{\p}{\p\sigma^0} + \frac{\p}{\p\sigma^1}\,.
\ee
This worldsheet theory \eqref{eq:NCstringflat} in flat spacetime is invariant under various global symmetry transformations of the worldsheet fields, which, in retrospective, already determines the spacetime symmetry algebra to be the extended string Galilei algebra \cite{nrGalilei, stringyNC}. This is analogous to relativistic string theory, in which global symmetries of the worldsheet theory in flat spacetime determine the Poincar\'{e} algebra to be the symmetry algebra of spacetime.

It is also possible to formulate nonrelativistic string theory in a Nambu-Goto-like formulation. Integrating out the worldsheet fields $\lambdar$ and $\lambdal$ in \eqref{eq:NCstring} yields the following two constraints:
\be
	\epsilon^{\alpha\beta} e_\alpha \p_\beta x^\mu \tlV_\mu= 0\,,
		\qquad
	\epsilon^{\alpha\beta} \bar e_\alpha \p_\beta x^\mu \bar \tlV_\mu = 0\,.
\ee
These two constraints imply that  $\wm_{\alpha\beta} = \tl_{\alpha\beta}\equiv \p_\alpha x^\mu \p_\beta x^\nu \tl_{\mu\nu}$ up to a conformal factor.
Plugging this solution into  the sigma model  action \eqref{eq:NCstring} we arrive at the following  Nambu-Goto-like formulation of nonrelativistic string theory (see also \cite{Gomis:2005pg,stringyNC}):
\begin{align} \label{eq:NGNCstring}
	S_\text{NG} & = - \frac{T}{2} \int d^2\sigma \lr \sqrt{-\tl} \, \tl^{\alpha\beta} \p_\alpha x^\mu \p_\beta x^\nu \ttr_{\mu\nu} + \epsilon^{\alpha\beta} \p_\alpha x^\mu \p_\beta x^\nu \tB_{\mu\nu} \rr \notag \\
		& \quad + \frac{1}{4\pi} \int d^2 \sigma \sqrt{-\tl} \, R(\tl) \, \Phi \,,
\end{align}
where $\tl \equiv \det \tl_{\alpha\beta}$ and $ \sqrt{-\tl} \, d^2\sigma$ defines the volume 2-form on $\Sigma$. Furthermore, $\tl^{\alpha\beta}$ is the inverse of the two by two matrix $\tl_{\alpha\beta}$. The Ricci scalar $R(\tl)$ is defined with respect to the pullback metric $\tl_{\alpha\beta}$.

We note that the nonrelativistic string sigma model defined in \eqref{eq:NCstring} and \eqref{eq:NGNCstring} trivializes if one reduces the target space tangent symmetry from the extended string Galilei algebra to the Bargmann algebra.\,\footnote{In the latter case  there is only one longitudinal timelike direction $A=0$, which leads to degenerate terms in \eqref{eq:NCstring} and \eqref{eq:NGNCstring}. To see explicitly that $S_\text{NG}$ is degenerate, we note that $\tl_{\mu\nu} = - \tlV_\mu{}^0 \tlV_\nu{}^0$ and thus $\tl = 0$ in the Bargmann case.} This selects the string Newton-Cartan geometry (associated with the extended string Galilei algebra) as the appropriate background structure for  nonrelativistic string theory, as opposed to Newton-Cartan geometry (associated with the Bargmann algebra).\,\footnote{See   footnote \ref{footnote:NC} for more details on Newton-Cartan geometry.} The string Newton-Cartan geometry is to nonrelativistic string theory what Riemannian geometry is to relativistic string theory.

In this paper we will exclusively work with the  Polyakov string action \eqref{eq:NCstring}.

\section{T-duality of Nonrelativistic String Theory}

Our next goal is to study the consequences of worldsheet duality acting on the path integral of the nonrelativistic   string  sigma model defined in  \eqref{eq:NCstring}. A nonrelativistic string   propagating on different backgrounds that are related
by  a duality transformation gives rise to the same physics. The backgrounds are related by a  T-duality transformation, which we derive by implementing the worldsheet duality transformation on the sigma model path integral.  Due to the  foliation structure of  the string Newton-Cartan geometry, there are three distinct types of duality transformations that can be implemented: one may transform along a spatial isometry direction  that is either longitudinal or transverse; moreover, for   completeness, one may also introduce a lightlike isometry in the longitudinal direction and   perform a T-duality transformation in this  lightlike  direction. We will study these three cases in turn.

\subsection{Longitudinal Spatial T-Duality} \label{sec:longflat}

We now assume that the string sigma model defined by \eqref{eq:NCstring} has a longitudinal spatial Killing vector $k^\mu$, i.e.
 \begin{equation}\label{longitudinalKilling}
 \tlV_\mu{}^0 k^\mu=0,\hskip 1truecm  \tlV_\mu{}^1 k^\mu\ne0\,,\hskip 1truecm \ttrV_\mu{}^{A'}k^{\mu}= 0\,.
 \end{equation}
We introduce a coordinate system $x^\mu = (y, x^{i})$
adapted to $k^\mu$,    such that $k^\mu \p_\mu = \p_y$\,. We note that $x^i$ contains a longitudinal coordinate.    Then, the associated abelian isometry is represented by a translation in the longitudinal spatial direction $y$. It is also possible to perform the duality transformation by gauging the isometry as in 
\cite{RocekVerlinde}. From  \eqref{longitudinalKilling}, it follows that
\begin{equation}
\tlV_y{}^0 =0, \ \ \ \tlV_y{}^1 \ne 0\,,\ \ \ \ttrV_y{}^{A^\prime}= 0 \hskip .5truecm \rightarrow \hskip .5truecm \tau_y = - \bar\tau_y \ne 0\,.
\ee
In this adapted coordinate system, all background fields and general coordinate transformation (g.c.t.)~parameters are  independent of $y$.

We perform a T-duality transformation along the isometry $y$-direction by  first defining
\be
	v_\alpha = \p_\alpha y\,.
\ee
The nonrelativistic string action   \eqref{eq:NCstring} is equivalent to the following ``parent"  action:
\begin{align} 
	S_\text{parent} & = -\frac{T}{2} \int d^2 \sigma \sqrt{-\wm} \, \wm^{\alpha\beta} \lr v_\alpha v_\beta \ttr_{yy} + 2 v_\alpha \p_\beta x^i \ttr_{yi} + \p_\alpha x^i \p_\beta x^j \ttr_{ij} \rr \notag \\
	& \quad - \frac{T}{2} \int d^2 \sigma \, \epsilon^{\alpha\beta} \ls \lambdar \, \wmV_\alpha \! \lr v_\beta \tlV_y + \p_\beta x^i \tlV_i \rr + \lambdal \, \bar{\wmV}_\alpha \! \lr v_\beta \overline{\tlV}_y + \p_\beta x^i \overline{\tlV}_i \rr \rs \notag \\
	& \quad - \frac{T}{2} \int d^2 \sigma \, \epsilon^{\alpha\beta} \lr 2 v_\alpha \p_\beta x^i \tB_{yi} + \p_\alpha x^i \p_\beta x^j \tB_{ij} + 2 \, \tilde{y} \, \p_\alpha v_\beta \rr \notag \\
	& \quad + \frac{1}{4\pi} \int d^2 \sigma \sqrt{-\wm} \, R\, \Phi \,.
\end{align}
In $S_\text{parent}$, $v_\alpha$ is considered to be an independent field. Moreover, $\tilde{y}$ is an  auxiliary field that plays the role of a Lagrange multiplier imposing the Bianchi identity $\epsilon^{\alpha\beta} \p_\alpha v_\beta = 0$. Obviously, solving this Bianchi identity leads us back to the original action \eqref{eq:NCstring}. Instead, we consider the equation of motion for $v_\alpha$,
\be
	\frac{\delta S_\text{parent}}{\delta v_\alpha} = 0\,,
\ee
which is solved by
\be\label{solution1}
	v_\alpha = - \frac{\ttr_{yi}}{\ttr_{yy}} \p_\alpha x^i + \frac{\wm_{\alpha\beta} \epsilon^{\beta\gamma}}{\ttr_{yy}\sqrt{-\wm}} \Bigl [ \tfrac{1}{2} \!\lr \lambdar \, \wmV_\gamma \tlV_y + \lambdal \, \bar{\wmV}_\gamma \bar\tlV_y \rr - \p_\gamma x^i \tB_{yi} - \p_\gamma \tilde{y} \, \Bigr ] \,.
\ee
Integrating out $v_\alpha$ by substituting the solution \eqref{solution1} back into $S_\text{parent}$,   we obtain the dual action
\begin{align} \label{eq:longdualprime}
	S'_\text{long.} & = - \frac{T}{2} \int d^2 \sigma \lr \sqrt{-\wm} \, \wm^{\alpha\beta} \p_\alpha \tilde{x}^\mu \p_\beta \tilde{x}^\nu \ttr'_{\mu\nu}  + \epsilon^{\alpha\beta} \p_\alpha \tilde{x}^\mu \p_\beta \tilde{x}^\nu \tB'_{\mu\nu}  \rr \notag \\
		& \quad - \frac{T}{2} \int d^2 \sigma \, \frac{1}{\ttr_{yy}} \lr \tl_{yy} \sqrt{-\wm} \, \lambdal \, \lambdar - \lambdar \, \zeta - \lambdal \, \bar{\zeta} \rr + \frac{1}{4\pi} \int d^2 \sigma \sqrt{-\wm} \, R\, \Phi'\,,
\end{align}
where $\tilde{x}^\mu = (\tilde{y}, x^i)$ and
\begin{subequations}
\begin{align}
	\zeta & = \sqrt{-\wm} \, \wm^{\alpha\beta} \wmV_\alpha \lr \p_\beta x^i \tB_{yi} + \p_\beta \tilde{y} \, \rr \tlV_y - \epsilon^{\alpha\beta} \, \wmV_\alpha \p_\beta x^i \lr \ttr_{yy} \tlV_i - \ttr_{yi} \tlV_y \rr\,, \\[2pt]
	\bar{\zeta} & = \sqrt{-\wm} \, \wm^{\alpha\beta} \bar{\wmV}_\alpha \lr \p_\beta x^i \tB_{yi} + \p_\beta \tilde{y} \, \rr \bar{\tlV}_y - \epsilon^{\alpha\beta} \, \bar{\wmV}_\alpha \p_\beta x^i \lr \ttr_{yy} \bar{\tlV}_i - \ttr_{yi} \bar{\tlV}_y \rr\,.
\end{align}
\end{subequations}
Moreover,
\begin{subequations} \label{eq:longBuscher}
\begin{align}
	\ttr'_{yy} & = \frac{1}{\ttr_{yy}}\,,
		&
	\Phi' & = \Phi - \frac{1}{2} \log \ttr_{yy}\,, \\[2pt]
	\ttr'_{yi} & = \frac{\tB_{yi}}{\ttr_{yy}}\,,
		&
	\tB'_{yi} & = \frac{\ttr_{yi}}{\ttr_{yy}}\,, \\[2pt]
	\ttr'_{ij} & = \ttr_{ij} + \frac{\tB_{yi} \tB_{yj} - \ttr_{yi} \ttr_{yj}}{\ttr_{yy}}\,,
		&
	\tB'_{ij} & = \tB_{ij} + \frac{\tB_{yi} \ttr_{yj} - \tB_{yj} \ttr_{yi}}{\ttr_{yy}}\,.
\end{align}
\end{subequations}
The shift of the dilaton $\Phi$ comes by regularizing as in \cite{Buscher:1987qj} the determinant in the path integral as the result of integrating out $v_\alpha$. The transformations \eqref{eq:longBuscher} are akin to  the Buscher rules \cite{Buscher:1987sk} in relativistic string theory.

In order to complete  the T-duality transformation we integrate out  $\lambdar$ and $\lambdal$, whose equations of motion are given by
\be \label{eq:sollambdapm}
	\lambdal = \frac{\zeta}{\tl_{yy} \sqrt{-\wm}}\,,
		\qquad
	\lambdar = \frac{\bar{\zeta}}{\tl_{yy} \sqrt{-\wm}}\,.
\ee
Substituting \eqref{eq:sollambdapm} back into $S'_\text{long.}$,  we find that the dual action takes the following equivalent form:
\begin{align} \label{eq:longdual}
	\tilde{S}_\text{long.} & = - \frac{T}{2} \int d^2 \sigma \lr \sqrt{-\wm} \, \wm^{\alpha\beta} \,  \p_\alpha \tilde{x}^\mu \p_\beta \tilde{x}^\nu \,\tilde{G}_{\mu\nu} + \epsilon^{\alpha\beta} \,  \p_\alpha \tilde{x}^\mu \p_\beta \tilde{x}^\nu\, \tilde{\tB}_{\mu\nu}\rr \notag \\
		& \quad + \frac{1}{4\pi} \int d^2 \sigma \sqrt{-\wm} \, R\, \tilde \Phi\,,
\end{align}
where $\tilde{x}^\mu = (\tilde{y}, x^i)$ and
\begin{subequations} \label{eq:longtrans}
\begin{align}
	\tilde{G}_{yy} & = 0\,,
		&
	& \hspace{-6.01cm}\tilde{\Phi} = \Phi - \frac{1}{2} \log \tl_{yy}\,, \label{eq:longtrans1}\\
	\tilde{G}_{yi} & = \frac{\tlV_i{}^A \tlV_y{}^B \epsilon_{AB}}{\tl_{yy}}\,,
		&
	& \hspace{-6.22cm}\tilde{\tB}_{yi} = \frac{\tl_{yi}}{\tl_{yy}}\,, \\[2pt]
	\tilde{G}_{ij} & = \ttr_{ij} + \frac{\lr \tB_{yi} \tlV_j{}^A + \tB_{yj} \tlV_i{}^A \rr \tlV_{y}{}^B \epsilon_{AB} + \ttr_{yy} \tl_{ij} - \ttr_{yi} \tl_{yj} - \ttr_{yj} \tl_{yi}}{\tl_{yy}}\,, \\
	\tilde{\tB}_{ij} & = \tB_{ij} + \frac{\tB_{yi} \tl_{yj} - \tB_{yj} \tl_{yi} - \lr \ttr_{yy} \tlV_i{}^A \tlV_j{}^B - \ttr_{yi} \tlV_y{}^A \tlV_{j}{}^B + \ttr_{yj} \tlV_y{}^A \tlV_{i}{}^B \rr \epsilon_{AB}}{\tl_{yy}}\,.
\end{align}
\end{subequations}
We note that integrating out $\lambdar$ and $\lambdal$ contributes   a determinant in the path integral, which can be regularized in the same way as it is done for the determinant originating from integrating out $v_\alpha$ \cite{Buscher:1987qj}. This determinant contributes a shift to the dilaton $\Phi'$, which leads to the following expression for the T-dual of $\Phi$:
\be
	\tilde{\Phi} = \Phi' - \frac{1}{2} \log \frac{\tl_{yy}}{\ttr_{yy}} = \Phi - \frac{1}{2} \log \tl_{yy}\,.
\ee
These T-duality transformations act in a very complicated way on the fundamental fields of the string Newton-Cartan geometry $\tlV_\mu{}^A\,, \ttrV_\mu{}^{A^\prime}$ and $m_\mu{}^A$ but much simpler on the string Galilei boost invariant variables $\tl_{\mu\nu}$ and $\ttr_{\mu\nu}$ we have introduced earlier.

Starting
 with  the action \eqref{eq:NCstring} that describes a nonrelativistic  string on a string Newton-Cartan background, which is not endowed with a Riemannian metric, we find that the T-dual action is given by    \eqref{eq:longdual}, which  is the action of a relativistic string propagating on a Lorentzian, Riemannian geometry with a lightlike isometry. The lightlike nature of the dual coordinate $\tilde{y}$ follows from the fact that $\tilde{G}_{yy} = 0$ in \eqref{eq:longtrans1}.

We note that, in \eqref{eq:longtrans}, a given general relativity  background is mapped under T-duality to many different string Newon-Cartan backgrounds.\,\footnote{We thank the referee for raising this question.} This is related to the fact that  the corresponding sigma model action for strings on these different string Newton-Cartan backgrounds are related to each other by the following field redefinitions of the Lagrange multipliers:\,\footnote{The rescaling factors in front of $\lambda''$ and $\bar{\lambda}''$ are taken to be the same so that there is no longitudinal Lorentz boost being introduced. This boost symmetry is already fixed by committing to a coordinate system adapted to the longitudinal isometry direction $y$\,.}
\begin{align} \label{eq:fieldred}
	\lambda = C \lambda'' + \frac{1}{\sqrt{-h}} \epsilon^{\alpha\beta} \, \bar{e}_\alpha \p_\beta x^\mu \, \bar{C}_\mu\,,
		\qquad
	\bar{\lambda} = C \, \bar{\lambda}'' + \frac{1}{\sqrt{-h}} \epsilon^{\alpha\beta} \, e_\alpha \p_\beta x^\mu \, C_\mu\,.
\end{align}
where   $C, C_\mu$ and ${\bar C}_\mu$ are arbitrary functions.
After these field redefinitions the non-relativistic string action \eqref{eq:NCstring} reads
\begin{align}
	S & = - \frac{T}{2} \int d^2 \sigma \ls \sqrt{-h} \, h^{\alpha\beta} \p_\alpha x^\mu \p_\beta x^\nu H''_{\mu\nu} + \epsilon^{\alpha\beta} \bigl( \lambda'' \, e_\alpha \p_\beta x^\mu \tau''_\mu + \bar\lambda'' \, \bar{e}_\alpha \p_\beta x^\mu \, \bar\tau''_\mu \bigr) \rs \notag \\
		& \quad - \frac{T}{2} \int d^2 \sigma \, \epsilon^{\alpha\beta} \p_\alpha x^\mu \p_\beta x^\nu B''_{\mu\nu} + \frac{1}{4\pi} \int d^2 \sigma \sqrt{-h} \, R \, \Phi''\,.
\end{align}
Here, with $C_\mu \equiv C_\mu{}^0 + C_\mu{}^1$ and $\bar{C}_\mu \equiv C_\mu{}^0 - C_\mu{}^1$\,, we have
\begin{align} \label{eq:fieldredsNC}
	H''_{\mu\nu} & = H_{\mu\nu} - \lr C_\mu{}^A \tau_\nu{}^B + C_\nu{}^A \tau_\mu{}^B \rr \eta_{AB} \,, &
	B''_{\mu\nu} & = B_{\mu\nu} + \lr C_\mu{}^A \tau_\nu{}^B - C_\nu{}^A \tau_\mu{}^B \rr \epsilon_{AB}\,, \notag \\[3pt]
	\tau''_\mu & = C \, \tau_\mu\,, \quad
	\bar\tau''_\mu = C \, \bar\tau_\mu\,, 
	& \Phi'' & = \Phi + \log C\,.
\end{align}
Plugging \eqref{eq:fieldredsNC} into \eqref{eq:longtrans} one can show that the $C$-function dependence drops out in the Buscher rules, as expected. By making special choices for the $C$-functions, one can always arrange it that, for instance,
$\tau''_{yy}$\,, $H_{y\mu}^{\prime\prime}$ and $B_{y\mu}^{\prime\prime}$ are fixed, in which case the remaining string Newton-Cartan data in \eqref{eq:longtrans} are uniquely determined for given $\tilde{G}_{\mu\nu}$ and $\tilde{B}_{\mu\nu}$\,.

Let us now discuss how to perform the inverse T-duality transformation to map the relativistic string action $\tilde{S}_\text{long.}$ in \eqref{eq:longdual} back to the nonrelativistic string action \eqref{eq:NCstring}. We start with defining $\tilde{v}_\alpha = \p_\alpha \tilde{y}$. Then, we define a parent action $\tilde{S}_\text{parent}$ that is equivalent to $\tilde{S}_\text{long.}$,
\be \label{eq:tparent}
	\tilde{S}_\text{parent} = \tilde{S}_\text{long.} (\p_\alpha \tilde{y} \rightarrow \tilde{v}_\alpha) - T \int d^2 \sigma \, \epsilon^{\alpha\beta} y \, \p_\alpha \tilde{v}_\beta\,,
\ee
where $\tilde{S}_\text{long.} (\p_\alpha \tilde{y} \rightarrow \tilde{v}_\alpha)$ is obtained by replacing $\p_\alpha \tilde{y}$ with $\tilde{v}_\alpha$ in \eqref{eq:longdual}. Moreover, $y$ is a Lagrange multiplier that imposes the Bianchi constraint $\epsilon^{\alpha\beta} \p_\alpha \tilde{v}_\beta = 0$. Solving this Bianchi identity leads us back to $\tilde{S}_\text{long.}$ in \eqref{eq:longdual}. Instead, we would like to integrate out $\tilde v_\alpha$ in the path integral to compute the dual action of $\tilde{S}_\text{long.}$. Note that, since $\tilde{G}_{yy} = 0$, $\tilde{S}_\text{parent}$ is linear in $\tilde v_\alpha$.

Before performing the $\tilde{v}_\alpha$ integral, let us use the dictionary in \eqref{eq:longtrans} to rewrite $\tilde{G}_{\mu\nu}$ and $\tilde{\tB}_{\mu\nu}$ in $\tilde{S}_\text{parent}$ in terms of the string Newton-Cartan data $\tlV_\mu{}^A$, $\ttr_{\mu\nu}$ and $B_{\mu\nu}$. Then, we introduce back the auxiliary fields $\lambda$ and $\bar{\lambda}$ and 
rewrite $\tilde{S}_\text{parent}$ as
\be
	S'_\text{parent} = S'_\text{long.} (\p_\alpha \tilde{y} \rightarrow \tilde{v}_\alpha) - T \int d^2 \sigma \, \epsilon^{\alpha\beta} y \, \p_\alpha \tilde{v}_\beta\,,
\ee
where $S'_\text{long.} (\p_\alpha \tilde{y} \rightarrow \tilde{v}_\alpha)$ is obtained by replacing $\p_\alpha \tilde{y}$ with $\tilde{v}_\alpha$ in \eqref{eq:longdualprime}.
Now, $S'_\text{parent}$ is quadratic in $\tilde{v}_\alpha$. Integrating out $\tilde{v}_\alpha$ in $S'_\text{parent}$ reproduces the nonrelativistic string action in \eqref{eq:NCstring}, including the appropriate dilaton field. Thus we conclude that the relativistic string action propagating on a Lorentzian, Riemannian background with a compact lightlike isometry can be mapped to the action \eqref{eq:NCstring} of a nonrelativistic string  moving in a  string Newton-Cartan background. We note that in order to define T-duality of relativistic string theory along a lightlike direction requires introducing additional worldsheet fields $\lambda$ and $\bar\lambda$, which goes beyond the well-known path integral manipulations considered by Buscher.

 As a particular case, we find that, for a  nonrelativistic string in flat spacetime, the T-dual along   a longitudinal spatial circle is given by a   relativistic string moving  in a flat Lorentzian spacetime with a lightlike compactified coordinate. This flat spacetime result was   anticipated by different means in \cite{Gomis:2000bd,Danielsson:2000gi}. In this way, we have established the relation between the DLCQ of relativistic string theory
 on an arbitrary  Lorentzian, Riemannian background and nonrelativistic string theory  on  the T-dual string Newton-Cartan background.\,\footnote{This relation was noticed for a particular curved background in \cite{Gomis:2005pg}.}

\subsection{Longitudinal Lightlike T-Duality} \label{sec:null}

We have shown in the previous subsection that the T-dual of relativistic string theory with a lightlike compactified circle is nonrelativistic string theory on a string Newton-Cartan background with a longitudinal spatial circle. It is then natural to ask a formal question: what happens if one T-dualizes the nonrelativistic string action \eqref{eq:NCstring} along a lightlike isometry direction? We will show in this subsection that a lightlike T-duality transformation maps   nonrelativistic string theory on a string Newton-Cartan background to nonrelativistic string theory on a T-dual string Newton-Cartan background with a longitudinal lightlike isometry. Here, the longitudinal lightlike T-duality is presented for   completeness, its physical significance is, however, not clear.

Let us assume that the string sigma model defined by \eqref{eq:NCstring} has a longitudinal lightlike Killing vector $\ell^\mu$ in the longitudinal sector, i.e.
 \begin{equation}\label{longitudinalnullKilling}
 \tlV_\mu \ell^\mu \ne 0\,,\hskip 1truecm  \bar{\tlV}_\mu \ell^\mu = 0\,,\hskip 1truecm \ttrV_\mu{}^{A^\prime} \ell^{\mu}= 0\,.
 \end{equation}
We define a coordinate system, $x^\mu = (u, x^{i})$,
adapted to $\ell^\mu$, such that $\ell^\mu \p_\mu = \p_u$.   Then, the associated abelian isometry is represented by a translation in the longitudinal lightlike direction $u$. From  \eqref{longitudinalnullKilling}, it follows that
\begin{equation}
\tlV_u \ne 0\,,\hskip 1truecm \bar{\tlV}_u = 0\,,\hskip 1truecm \ttrV_u{}^{A^\prime}= 0\,.
\end{equation}
In this adapted coordinate system, all background fields and g.c.t.~parameters are  independent of $u$.

To perform a T-duality transformation along the lightlike isometry $u$-direction, it is convenient to introduce an auxiliary field $f_\alpha$.
Then, we rewrite the sigma model of nonrelativistic string theory \eqref{eq:NCstring} as
\begin{align} \label{eq:NCstringnull}
	S_\text{light.} & = -\frac{T}{2} \int \! d^2 \sigma \lr \sqrt{-\wm} \, \wm^{\alpha\beta} \p_\alpha x^\mu \p_\beta x^\nu \ttr_{\mu\nu} + \epsilon^{\alpha\beta} \p_\alpha x^\mu \p_\beta x^\nu \tB_{\mu\nu} \rr + \frac{1}{4\pi} \int \! d^2 \sigma \sqrt{-\wm} \, R\, \Phi \notag \\
		& \quad - \frac{T}{2} \int \! d^2 \sigma \, \epsilon^{\alpha\beta} \lr \wmV_\alpha f_\beta + 2 \zetar \, f_\alpha \p_\beta x^\mu \tlV_\mu + \lambdal \, \bar{\wmV}_\alpha \p_\beta x^i \bar{\tlV}_i \rr\,,
\end{align}
where $\zetar$ is a Lagrange multiplier that imposes a constraint,
\be
	\epsilon^{\alpha\beta} f_\alpha \p_\beta x^\mu \tlV_\mu = 0\,.
\ee
Integrating out $\zetar$ sets 
\be \label{eq:solf}
	f_{\alpha} = \lambdar \, \p_{\alpha} x^\mu \tlV_\mu\,.
\ee 
Plugging this solution into $S_\text{light.}$ to eliminate $f_\alpha$ we reproduce the sigma model of nonrelativistic string theory \eqref{eq:NCstring} with $\bar\tau_u = 0$. Note that the worldsheet field $\lambda$ reappears in the solution to $f_\alpha$ as an integration constant. Next, let us define
\be
	v_\alpha = \p_\alpha u\,.
\ee
Then, $S_\text{light.}$ is equivalent to the following parent action:
\begin{align} \label{eq:longparent}
	S_\text{parent} & = -\frac{T}{2} \int d^2 \sigma \sqrt{-\wm} \, \wm^{\alpha\beta} \lr  v_\alpha v_\beta \ttr_{uu} + 2 v_\alpha \p_\beta x^i \ttr_{ui} + \p_\alpha x^i \p_\beta x^j \ttr_{ij} \rr \notag \\
	& \quad - \frac{T}{2} \int d^2 \sigma \, \epsilon^{\alpha\beta} \ls \wmV_\alpha f_\beta  + 2 \zetar f_\alpha \lr v_\beta \tlV_u + \p_\beta x^i \tlV_i \rr + \lambdal \, \bar{\wmV}_\alpha \p_\beta x^i \bar{\tlV}_i \rs \notag \\
	& \quad - \frac{T}{2} \int d^2 \sigma \, \epsilon^{\alpha\beta} \lr 2 v_\alpha \p_\beta x^i \tB_{ui} + \p_\alpha x^i \p_\beta x^j \tB_{ij} + 2 \tilde{u} \, \p_\alpha v_\beta \rr \notag \\
	& \quad + \frac{1}{4\pi} \int d^2 \sigma \sqrt{-\wm} \, R\, \Phi \,.
\end{align}
In $S_\text{parent}$\,, $v_\alpha$ is considered to be an independent field. Moreover, $\tilde{u}$ is an  auxiliary field that plays the role of a Lagrange multiplier imposing the Bianchi identity $\epsilon^{\alpha\beta} \p_\alpha v_\beta = 0$. Obviously, solving this Bianchi identity leads us back to $S_\text{light.}$. Instead, we consider the equation of motion for $v_\alpha$,
\be
	\frac{\delta S_\text{parent}}{\delta v_\alpha} = 0\,,
\ee
which is solved by
\be
	v_\alpha = - \frac{\ttr_{ui}}{\ttr_{uu}} \p_\alpha x^i + \frac{\wm_{\alpha\beta} \epsilon^{\beta\gamma}}{\ttr_{uu}\sqrt{-\wm}} \Bigl( \zetar f_\gamma \tlV_u - \p_\gamma x^i \tB_{ui} - \p_\gamma \tilde{u}\Bigr)\,.
\ee
If we integrate out $v_\alpha$ by substituting this solution back into $S_\text{parent}$, then we obtain the dual action,
\begin{align} \label{eq:nulldualprime}
	S'_\text{light.} & = - \frac{T}{2} \int d^2 \sigma \lr \sqrt{-\wm} \, \wm^{\alpha\beta} \p_\alpha \tilde{x}^\mu \p_\beta \tilde{x}^\nu \ttr'_{\mu\nu} + \epsilon^{\alpha\beta} \p_\alpha x^\mu \p_\beta x^\nu \tB'_{\mu\nu} \rr \notag \\[3pt]
		& \quad - \frac{T}{2} \int d^2 \sigma \, \epsilon^{\alpha\beta} \lr \wmV_\alpha f_\beta + \lambdal \, \bar{\wmV}_\alpha \p_\beta x^i \bar{\tlV}_i \rr + \frac{1}{4\pi} \int d^2 \sigma \sqrt{-\wm} \, R\, \Phi' \, \notag \\[3pt]
		& \quad - \frac{T}{2} \int d^2 \sigma \, \frac{1}{\ttr_{uu}} \Bigl[ \lr \zetar \, \tlV_u \rr^2 \! \sqrt{-\wm} \, \wm^{\alpha\beta} f_\alpha f_\beta - 2 \zetar \, \wm^{\alpha\beta} f_\alpha \, \xi_\beta \Bigr]\,,
\end{align}
where $\tilde{x}^\mu = (\tilde{u}, x^i)$ and
\begin{align}
	\xi_\alpha & = \sqrt{-\wm} \lr \p_\alpha x^i \tB_{ui} + \p_\alpha \tilde{u} \rr \tlV_u - \wm_{\alpha\beta} \epsilon^{\beta\gamma} \p_\gamma x^i \lr \ttr_{uu} \tlV_i - \ttr_{ui} \tlV_u \rr\,.
\end{align}
Moreover,
\begin{subequations} \label{eq:lighttransBuscher}
\begin{align}
	\ttr'_{uu} & = \frac{1}{\ttr_{uu}}\,,
		&
	\Phi' & = \Phi - \frac{1}{2} \log \ttr_{uu}\,, \\[2pt]
	\ttr'_{ui} & = \frac{\tB_{ui}}{\ttr_{uu}}\,,
		&
	\tB'_{ui} & = \frac{\ttr_{ui}}{\ttr_{uu}}\,, \\[2pt]
	\ttr'_{ij} & = \ttr_{ij} + \frac{\tB_{ui} \tB_{uj} - \ttr_{ui} \ttr_{uj}}{\ttr_{uu}}\,,
		&
	\tB'_{ij} & = \tB_{ij} + \frac{\tB_{ui} \ttr_{uj} - \tB_{uj} \ttr_{ui}}{\ttr_{uu}}\,.
\end{align}
\end{subequations}
The shift of the dilaton $\Phi$ comes by regularizing as in \cite{Buscher:1987qj} the determinant in the path integral from integrating out $v_\alpha$.

In order to complete the T-duality transformation, we integrate out $f_\alpha$ in $S'_\text{light.}$, whose equation of motion is
\be \label{eq:solHm}
	f_\alpha = \frac{\ttr_{uu} \wm_{\alpha\beta} \, \epsilon^{\beta\gamma} \wmV_\gamma + 2 \zetar \, \xi_\alpha}{2 \lr \zetar \, \tlV_u\rr^2 \sqrt{-\wm}}\,.
\ee
Substituting \eqref{eq:solHm} back into $S'_\text{light.}$, the dual action takes the following equivalent form:
\begin{align} \label{eq:lightdual}
	\tilde{S}_\text{light.} & = -\frac{T}{2} \int d^2 \sigma \ls \sqrt{-\wm} \, \wm^{\alpha\beta} \, \p_\alpha \tilde{x}^\mu \p_\beta \tilde{x}^\nu \tilde{\ttr}_{\mu\nu} + \epsilon^{\alpha\beta} \lr \lambdar \, \wmV_\alpha \p_\beta \tilde{x}^\mu \tilde{\tlV}_\mu + \lambdal \, \bar \wmV_\alpha \p_\beta x^i \, \bar \tlV_i \rr \rs \notag \\
		& \quad - \frac{T}{2} \int d^2 \sigma \, \epsilon^{\alpha\beta} \p_\alpha \tilde{x}^\mu \p_\beta \tilde{x}^\nu \tilde{\tB}_{\mu\nu} + \frac{1}{4\pi} \int d^2 \sigma \sqrt{-\wm} \, R\, \tilde{\Phi} \,,
\end{align}
where $\tilde{x}^\mu = (\tilde{u}, x^i)$ and
\begin{subequations} \label{eq:nulltrans}
\begin{align}
	\tilde{\tlV}_u & = \frac{1}{\tlV_u}\,, \\[2pt]
	\tilde{\tlV}_i & = \frac{\tB_{ui} \tlV_u - \ttr_{uu} \tlV_i + \ttr_{ui} \tlV_u}{\tlV_u \tlV_u}\,,
		&
	 \tilde{\Phi} & = \Phi - \log |\tlV_u|\,, \\[2pt]
	\tilde{\ttr}_{u\mu} & = 0\,,
		&
	\tilde{\tB}_{ui} & = \frac{\tlV_i}{\tlV_u}\,, \\[2pt]
	\tilde{\ttr}_{ij} & = \ttr_{ij} + \frac{\ttr_{uu} \tlV_{i} \tlV_{j} - \lr \ttr_{ui} \tlV_j + \ttr_{uj} \tlV_i \rr \tlV_{u}}{\tlV_{u} \tlV_{u}}\,,
		&
	\tilde{\tB}_{ij} & = \tB_{ij} + \frac{\tB_{ui} \tlV_{j} - \tB_{uj} \tlV_{i}}{\tlV_u}\,.
\end{align}
\end{subequations}
Note that $\bar\tlV_i$ remains unchanged. Moreover,
\be \label{eq:zetatolambda}
	\lambda = \frac{1}{\zetar}.
\ee
One may check that $\lambda$ and $\eta^{-1}$ indeed transform in the same way under worldsheet diffeomorphisms and worldsheet Weyl transformation. Note that integrating out $f_\alpha$ in $S'_\text{light.}$ contributes a determinant in the path integral, which can be regularized in the same way as it is done for the determinant from integrating out $v_\alpha$ \cite{Buscher:1987qj}. Moreover, the change of variables in \eqref{eq:zetatolambda} also contributes a Jacobian in the path integral, which cancels the $\eta$ dependence in the determinant from integrating out $f_\alpha$. Finally, these measure terms generate a shift to the dilaton $\Phi'$,
\be \label{eq:etatolambda}
	\tilde{\Phi} = \Phi' - \frac{1}{2} \log \frac{\tlV_u \tlV_u}{\ttr_{uu}} = \Phi - \log |\tlV_u| \,.
\ee

If one applies the duality transformations in \eqref{eq:nulltrans} again on $\tilde{\tlV}_\mu$, $\tilde{\ttr}_{\mu\nu}$ and $\tilde{\tB}_{\mu\nu}$, it does \emph{not} give back the original geometry $\tlV_\mu$, $\ttr_{\mu\nu}$ and $\tB_{\mu\nu}$. Nevertheless, the $\mathbb{Z}_2$ symmetry of the T-duality transformation is still preserved once we take into account the following field redefinition:
\be
	f_\alpha \rightarrow f_\alpha + \frac{e_\alpha}{2 \eta \sqrt{-h}} \, \epsilon^{\beta\gamma} \, \bar{e}_\beta \p_\gamma x^\mu \, \bar{C}_\mu\,.
\ee
This field redefinition gives rise in \eqref{eq:NCstringnull} to the following shifts of $H_{\mu\nu}$ and $B_{\mu\nu}$\,:
\be \label{eq:ufieldred}
	H_{\mu\nu} \rightarrow H_{\mu\nu} + \frac{1}{2} \lr \tau_\mu \bar{C}_\nu + \tau_\nu \bar{C}_\mu \rr\,,
		\qquad
	B_{\mu\nu} \rightarrow B_{\mu\nu} - \frac{1}{2} \lr \tau_\mu \bar{C}_\nu - \tau_\nu \bar{C}_\mu \rr\,.
\ee
Plugging \eqref{eq:ufieldred} back into \eqref{eq:nulltrans} one can show that $\bar{C}_\mu$ drops out in the Buscher rules, as expected.
By making special choices of the $\bar{C}_\mu$, 
one can always arrange it that $\ttr_{u\mu} = 0$\,. The T-duality rules are then given by
\begin{subequations} \label{eq:nullBuscher}
\begin{align}
	\tilde{\tlV}_u & = \frac{1}{\tlV_u}\,,
		&	
	 \tilde{\Phi} & = \Phi - \log |\tlV_u| \,, \label{eq:nullBuscher1} \\[2pt]
	\tilde{\tlV}_i & = \frac{\tB_{ui}}{\tlV_u}\,,
	 	&
	\tilde{\tB}_{ui} & = \frac{\tlV_i}{\tlV_u}\,, \\[2pt]
	\tilde{\ttr}_{ij} & = \ttr_{ij}\,,
		&
	\tilde{\tB}_{ij} & = \tB_{ij} + \frac{\tB_{ui} \tlV_{j} - \tB_{uj} \tlV_{i}}{\tlV_u}\,.
\end{align}
\end{subequations}
Note that $\tilde{H}_{u\mu} = 0$ remains unchanged. It is straightforward to check that applying the duality transformations \eqref{eq:nullBuscher} a second time indeed brings $\tilde{\tlV}_\mu{}$, $\tilde{\ttr}_{\mu\nu}$ and $\tilde{\tB}_{\mu\nu}$ back to the original fields  $\tlV_\mu$, $\ttr_{\mu\nu}$ and $\tB_{\mu\nu}$.

We could also have imposed the condition $\ttr_{u\mu}=0$ at the very beginning without affecting the final result for the T-duality rules. In fact, the procedure of the T-duality transformation simplifies significantly. Now, the parent action in \eqref{eq:longparent} becomes
\begin{align} \label{eq:Sparentadapted}
	S_\text{parent} & = -\frac{T}{2} \int d^2 \sigma \sqrt{-\wm} \, \wm^{\alpha\beta} \p_\alpha x^i \p_\beta x^j \ttr_{ij} + \frac{1}{4\pi} \int d^2 \sigma \sqrt{-\wm} \, R\, \Phi \notag \\
	& \quad - \frac{T}{2} \int d^2 \sigma \, \epsilon^{\alpha\beta} \ls \wmV_\alpha f_\beta  + 2 \zetar f_\alpha \lr v_\beta \tlV_u + \p_\beta x^i \tlV_i \rr + \lambdal \, \bar{\wmV}_\alpha \p_\beta x^i \bar{\tlV}_i \rs \notag \\
	& \quad - \frac{T}{2} \int d^2 \sigma \, \epsilon^{\alpha\beta} \lr 2 v_\alpha \p_\beta x^i \tB_{ui} + \p_\alpha x^i \p_\beta x^j \tB_{ij} + 2 \tilde{u} \, \p_\alpha v_\beta \rr \,,
\end{align}
which is linear in $v_\alpha$\,. Integrating out $v_\alpha$ in the path integral results in the following constraint on $f_\alpha$\,,
\be \label{eq:covariables}
	f_\alpha = \frac{1}{\zetar \tlV_u} \lr \p_\alpha x^i \tB_{ui} + \p_\alpha \tilde{u} \rr.
\ee
Plugging this solution to $f_\alpha$ back into \eqref{eq:Sparentadapted} and applying the change of variables in \eqref{eq:zetatolambda} reproduces the dual action $\tilde{S}_\text{light.}$ in \eqref{eq:lightdual} with $\tilde{\ttr}_{u\mu} = 0$ and the same $\tilde{H}_{ij}$, $\tilde{\tlV}_\mu$ and $\tilde{\tB}_{\mu\nu}$ as given in \eqref{eq:nullBuscher}. The shift in the dilaton field now comes from imposing the constraint on $f_\alpha$ in \eqref{eq:covariables}. In contrast, in the more involved procedure presented without fixing $\ttr_{u\mu}$ to zero, the shift of $\Phi$ can be derived in the standard way as in \cite{Buscher:1987qj}.\,\footnote{One may also use the field redefinitions in \eqref{eq:fieldred} to fix $\ttr_{y\mu}$ to zero in \S\ref{sec:longflat} to derive the T-duality transformation rules in \eqref{eq:longtrans} except for the dilaton shift.}

 We conclude that the T-duality transformation along a lightlike isometry direction maps to each other   nonrelativistic string theory on two different string Newton-Cartan background geometries, whose relations are given in \eqref{eq:nullBuscher}. In particular, this duality  maps between two lightlike circles of reciprocal radii.

\subsection{Transverse T-Duality} \label{sec:trans}

Finally, we consider the nonrelativistic string sigma model defined by \eqref{eq:NCstring} with  a transverse spatial Killing vector $p^\mu$, i.e.
 \begin{equation}\label{transverseKilling}
 \tlV_\mu{}^A p^\mu =0\,,\hskip 1truecm E_\mu{}^{A'}p^{\mu} \ne 0\,.
 \end{equation}
We define a coordinate system $x^\mu = (x^{i}, z)$
adapted to $p^\mu$, such that $p^\mu \p_\mu = \p_z$. Then, the associated abelian isometry is represented by a translation in the transverse direction $z$.
From  \eqref{transverseKilling}, it follows that
\begin{equation}
\tlV_z{}^A=0\,,\ \ \ \ttrV_z{}^{A^\prime}\ne 0\hskip .5truecm \rightarrow \hskip .5truecm \tlV_z = \bar\tlV_z = 0,\ \ \  \ttr_{zz}\ne 0\,.
 \end{equation}
In this adapted coordinate system, all background fields and g.c.t.~parameters are assumed to be independent of $z$. Under the above conditions, the string action \eqref{eq:NCstring} reduces to the following form:
\begin{align} \label{eq:NCstringtrans}
	S_\text{trans.} & = -\frac{T}{2} \int d^2 \sigma \sqrt{-\wm} \, \wm^{\alpha\beta} \lr \p_\alpha z \, \p_\beta z \, \ttr_{zz} + 2 \, \p_\alpha z \, \p_\beta x^i \, \ttr_{zi} + \p_\alpha x^i \p_\beta x^j \ttr_{ij} \rr \notag \\
		& \quad - \frac{T}{2} \int d^2 \sigma \, \epsilon^{\alpha\beta} \lr 2 \, \p_\alpha z \, \p_\beta x^i \tB_{zi} + \p_\alpha x^i \p_\beta x^j \tB_{ij} \rr + \frac{1}{4\pi} \int d^2 \sigma \sqrt{-\wm} \, R \, \Phi \, \notag \\
		& \quad - \frac{T}{2} \int d^2 \sigma \, \epsilon^{\alpha\beta} \lr \lambdar \, \wmV_\alpha \tlV_i + \lambdal \, \bar{\wmV}_\alpha \bar{\tlV}_i \rr \p_\beta x^i\,.
\end{align}
Remarkably, as far as the derivation of the T-duality rules is concerned, the $\lambda$ and $\bar\lambda$ terms in the last line of \eqref{eq:NCstringtrans} are not involved and the nonrelativistic string action is in form the same as the relativistic string action in the Polyakov formalism. Therefore, the dual action $\tilde{S}_\text{trans.}$ must take the same form as $S_\text{trans.}$\,,
\begin{align} \label{eq:NCstringtransdual}
	\tilde{S}_\text{trans.} & = -\frac{T}{2} \int d^2 \sigma \ls \sqrt{-\wm} \, \wm^{\alpha\beta} \, \p_\alpha \tilde{x}^\mu \p_\beta \tilde{x}^\nu \tilde{\ttr}_{\mu\nu} + \epsilon^{\alpha\beta} \lr \lambdar \, \wmV_\alpha \, \p_\beta x^\mu \, \tlV_\mu + \lambdal \, \bar \wmV_\alpha \, \p_\beta \tilde{x}^\mu \, \bar \tlV_\mu \rr \rs \notag \\
		& \quad - \frac{T}{2} \int d^2 \sigma \, \epsilon^{\alpha\beta} \p_\alpha \tilde{x}^\mu \p_\beta \tilde{x}^\nu \tilde{\tB}_{\mu\nu} + \frac{1}{4\pi} \int d^2 \sigma \sqrt{-\wm} \, R\, \tilde{\Phi} \,,
\end{align}
where $\tilde{x} = (x^i, z)$ and the transformations of various fields satisfy the T-duality rules,
\begin{subequations} \label{eq:transBuscher}
\begin{align}
	\tilde{\ttr}_{zz} & = \frac{1}{\ttr_{zz}}\,,
		&
	\tilde{\Phi} & = \Phi - \frac{1}{2} \log \ttr_{zz}\,, \\[2pt]
	\tilde{\ttr}_{zi} & = \frac{\tB_{zi}}{\ttr_{zz}}\,,
		&
	\tilde{\tB}_{zi} & = \frac{\ttr_{zi}}{\ttr_{zz}}\,, \\[2pt]
	\tilde{\ttr}_{ij} & = \ttr_{ij} + \frac{\tB_{zi} \tB_{zj} - \ttr_{zi} \ttr_{zj}}{\ttr_{zz}}\,,
		&
	\tilde{\tB}_{ij} & = \tB_{ij} + \frac{\tB_{zi} \ttr_{zj} - \tB_{zj} \ttr_{zi}}{\ttr_{zz}}\,.
\end{align}
\end{subequations}
The background fields $\tlV_\mu$ and $\bar\tlV_\mu$ remain unchanged. Performing the duality transformations \eqref{eq:transBuscher} again maps $\tilde{\ttr}_{\mu\nu}$ and $\tilde{\tB}_{\mu\nu}$ back to the original fields $\ttr_{\mu\nu}$ and $\tB_{\mu\nu}$. We thus obtain a duality  between nonrelativistic string theory  propagating in two different  string Newton-Cartan backgrounds with Kalb-Ramond and dilaton fields.
The difference between the  lightlike  T-duality rules \eqref{eq:nullBuscher} is that, for the transverse case, the duality transformations mix up the Kalb-Ramond field $\tB_{\mu\nu}$ with the transverse two-tensor $\ttr_{\mu\nu}$ instead of the longitudinal Vielbein $\tlV_{\mu}$. This transverse duality maps between two transverse circles of reciprocal radii.

\section{Conclusions} \label{sec:concl}

Nonrelativistic string theory is a   theory with rather distinctive features both in the worldsheet and in the target space in comparison to relativistic string theory. The degrees of freedom on the worldsheet go beyond the usual worldsheet fields parametrizing spacetime coordinates. The additional $\lambda$ and $\bar\lambda$ fields play a central role in the inner workings of nonrelativistic string theory. They are responsible for realizing the nonrelativistic spacetime symmetries on the worldsheet fields and endow nonrelativistic string theory with its distinctive string perturbation theory  \cite{Gomis:2000bd}.

Nonrelativistic strings couple to a very specific background geometric structure: string Newton-Cartan geometry. This geometry is ultimately dictated by the vertex operators of nonrelativistic string theory and is rather different from the familiar Riemannian geometry that relativistic strings couple to. The couplings of nonrelativistic string theory to an arbitrary string Newton-Cartan geometry are encoded in the  nonlinear sigma model \eqref{eq:NCstring}. String Newton-Cartan geometry is to nonrelativistic string theory what Riemannian geometry is to relativistic string theory. It would be interesting to write down the   sigma model for nonrelativistic superstring theory and investigate the corresponding superspace target space geometry.

We have studied   duality transformations of the path integral of the nonrelativistic string sigma model and derived an equivalence between string theories propagating in distinct, but T-dual backgrounds. The most interesting case is the action of T-duality along a longitudinal (spatial) direction. We have shown that nonrelativistic string theory coupled to a  string Newton-Cartan background  with a compact longitudinal spatial direction is equivalent to relativistic string theory propagating on a Lorentzian, Riemannian geometry with a compact lightlike isometry. This duality provides a tantalizing example of how string theory in a conventional geometric background (a Lorentzian geometry) is equivalent to string theory  with a non-Riemannian, but  still  recognizable geometric structure --- string Newton-Cartan geometry.

This general relation between nonrelativistic string theory  and   relativistic string theory with a lightlike compact isometry  provides a
  first principles definition of the worldsheet theory of relativistic string theory with a compact lightlike isometry, i.e. a definition of
 DLCQ\,\footnote{We recall that DLCQ stands for discrete light cone quantization.} of relativistic string theory. Until hitherto,  the DLCQ of relativistic string theory could only be defined by considering  a subtle, singular  infinite boost limit of a small spacelike circle \cite{Seiberg:1997ad,Sen:1997we,Hellerman:1997yu}. Instead, the nonrelativistic string theory sigma model gives a finite, explicit definition of DLCQ of relativistic string theory on an arbitrary Lorentzian, Riemannian metric with a lightlike isometry.  A key ingredient in defining DLCQ of relativistic string theory is the presence of the additional worldsheet fields $\lambda$ and $\bar{\lambda}$, that have no direct spacetime interpretation.
 The DLCQ of  string/M-theory has played a central role in various nonperturbative approaches, most notably in Matrix theory~\cite{Banks:1996vh,Susskind:1997cw,Seiberg:1997ad,Sen:1997we}. It would be interesting to use the worldsheet definition of the DLCQ of string theory
 on an arbitrary background to give a nonperturbative Matrix theory definition of string theory for a broader class of backgrounds and also to compute string amplitudes in DLCQ of relativistic string  theory using \eqref{eq:NCstring}, as was done for flat spacetime in \cite{Gomis:2000bd}. The study of boundary conditions in the nonrelativistic sigma model  and the effective field theory living on the corresponding D-branes provides a strategy to address this problem.

  We have also studied the   duality transformations of the path integral of the nonrelativistic string sigma model in a string Newton-Cartan background with a   longitudinal lightlike and a  transverse spatial  direction. We have shown that T-duality mixes the Kalb-Ramond field $\tB_{\mu\nu}$ with the longitudinal Vielbein $\tlV_{\mu}$ in the former case and  with the transverse two-tensor $\ttr_{\mu\nu}$ in the latter case. In both cases, however, in contrast to the duality transformation along a longitudinal spatial isometry direction, the T-dual theory remains a nonrelativistic string theory on a  string Newton-Cartan geometry.

Recently, there has been  work on general relativity with a  lightlike  isometry direction in the context of nonrelativistic strings \cite{HHO,Kluson}, where a ``null reduction" is applied to a relativistic string in order to obtain a string in a nonrelativistic background.\,\footnote{The background geometry discussed in \cite{HHO} can be viewed as a specialization of string Newton-Cartan geometry when there is no torsion. See more in footnote \ref{footnote:TNC}.} There is other recent work where a  particle limit of relativistic strings is considered leading to so-called Galilean strings with nonrelativistic worldsheets moving in a Newtonian spacetime
\cite{Batlle:2016iel,Gomis:2016zur,Batlle:2017cfa}; these different works  deal with  strings moving in a Newton-Cartan background with a one-dimensional foliation as opposed to the string Newton-Cartan background with a two-dimensional foliation that we consider in the current work. If one wishes to consider a nonrelativistic  theory with a   non-empty Hilbert space of string excitations,  one is led to consider the string Newton-Cartan geometry.
 There are also interesting connections with \cite{nrDFT, Morand:2017fnv}, where  nonrelativistic string theory in flat space \cite{Gomis:2000bd} is embedded  in the double field theory formalism.

Many interesting lines of investigation in nonrelativistic string theory remain, and we close with a few of them. The sigma model of nonrelativistic string theory is classically Weyl invariant and quantum consistency of the worldsheet theory determines the backgrounds on which nonrelativistic string theory can be consistently defined. It would be interesting to derive the spacetime equations of motion for the string Newton-Cartan fields (possibly including the foliation constraint \eqref{foliation}), the Kalb-Ramond field and the dilaton that determine the classical solutions of nonrelativistic string theory by analyzing the Weyl invariance of the worldsheet theory at the quantum level. It would also be interesting to derive the spacetime (string) field theory that
reproduces the S-matrix defined by the worldsheet correlation functions of nonrelativistic string theory. Last but not least, there are potential interesting applications to non-relativistic holography that are worth exploring.

\acknowledgments

 We~would like to thank Joaquim Gomis, Troels Harmark, Jelle Hartong, Niels Obers, Lorenzo Di Pietro, Jan Rosseel and Ceyda \c{S}im\c{s}ek for useful discussions. E.B.   thanks the Perimeter Institute for financial support and for providing a hospitable and stimulating research atmosphere.
Z.Y\!. thanks the Niels Bohr Institute and the University of Groningen for hospitality.
This research was supported in part by Perimeter Institute for Theoretical Physics. Research at Perimeter Institute is supported by the Government of Canada through the Department of Innovation, Science and Economic Development and by the Province of Ontario through the Ministry of Research, Innovation and Science.
\newpage
\bibliographystyle{JHEP}
\bibliography{nrtd}

\end{document}